\input amstex
\magnification 1200
\TagsOnRight
\def\qed{\ifhmode\unskip\nobreak\fi\ifmmode\ifinner\else
 \hskip5pt\fi\fi\hbox{\hskip5pt\vrule width4pt
 height6pt depth1.5pt\hskip1pt}}
 \def\adots{\mathinner{\mkern2mu\raise1pt\hbox{.}
\mkern3mu\raise4pt\hbox{.}\mkern1mu\raise7pt\hbox{.}}}
\def\sdots{\mathinner{
     \mskip.01mu\raise1pt\vbox{\kern1pt\hbox{.}}
     \mskip.01mu\raise3pt\hbox{.}
     \mskip.01mu\raise5pt\hbox{.}
\mskip1mu}}
\baselineskip 20 pt
\parskip 5 pt

\centerline {\bf A UNIFIED APPROACH TO DARBOUX TRANSFORMATIONS}

\vskip 10 pt
\centerline {Tuncay Aktosun}
\vskip -6 pt
\centerline {Department of Mathematics}
\vskip -6 pt
\centerline {University of Texas at Arlington}
\vskip -6 pt
\centerline {Arlington, TX 76019-0408, USA}

\centerline{Cornelis van der Mee}
\vskip -6 pt
\centerline{Dipartimento di Matematica e Informatica}
\vskip -6 pt
\centerline{Universit\`a di Cagliari}
\vskip -6 pt
\centerline{Viale Merello 92, 09123 Cagliari, Italy}

\vskip 10 pt

\noindent {\bf Abstract}: We analyze a certain class of integral equations
related to Marchenko equations and Gel'fand-Levitan equations associated with various
systems of ordinary differential operators. When the integral operator
is perturbed by a finite-rank perturbation, we explicitly evaluate
the change in the solution. We show how this result provides a unified
approach to Darboux transformations associated with various
systems of ordinary differential operators. We illustrate our
theory by deriving the Darboux transformation for the Zakharov-Shabat system
and show how the potential and wave function change when a discrete eigenvalue is added
to the spectrum.

\vskip 15 pt
\par \noindent {\bf Mathematics Subject Classification (2000):}
37K35 (34L40 35Q53 35Q55 37K15)
\vskip -6 pt
\par\noindent {\bf Keywords:}
Darboux transformation, Marchenko equation, Gel'fand-Levitan equation,
Zakharov-Shabat system, invariance principle, nonlinear Schr\"odinger equation

\vskip -6 pt
\par\noindent {\bf Short title:} A unified approach to Darboux transformations
\newpage

\noindent {\bf 1. INTRODUCTION}
\vskip 3 pt

Consider the one-parameter family of integral equations of the second kind
$$\beta(x,y)+\zeta(x,y)+\int_x^\infty dz\,\beta(x,z)\,\omega(z,y)=0,\qquad y>x,
\tag 1.1$$
where $\beta(x,y)$ is the unknown, $\zeta(x,y)$ is the nonhomogeneous term, and
$\omega(z,y)$ is an integral kernel which does not depend on the parameter
$x\in{\bold R}$ and satisfies
$$\sup_{y>x}\,\int_x^\infty dz\left(\|\omega(z,y)\|+\|\omega(y,z)\|\right)
<+\infty,\tag 1.2$$
where $||\cdot||$ denotes any $N\times N$-matrix norm.
Let us write (1.1) as
$$\beta+\zeta+\beta\Omega=0,\tag 1.3$$
where the integral operator $\Omega$ acts from the right. From
(1.2), as shown in the Appendix, it follows that $\Omega$ is bounded
on the complex Banach spaces ${\Cal H}_p^{M\times N}$ of
$M\times N$ matrix-valued measurable functions
$F:(x,+\infty)\to\bold C^{M\times N}$ such that the matrix norm
$\|F(\cdot)\|$
belongs to
$L^p(x,+\infty)$ for
$1\le p\le+\infty.$

We assume that,
for each $x\in{\bold R}$, $(I+\Omega)$ is an invertible operator on
${\Cal H}_1^{N\times N}$ and on ${\Cal H}_2^{N\times N},$
where $I$ denotes the identity operator.
Using $(I+R)$ to denote the corresponding resolvent operator, where
$$R:=(I+\Omega)^{-1}-I,\quad I+R=(I+\Omega)^{-1},\tag 1.4$$
the solution to (1.3)
can formally be written as
$$\beta=-\zeta(I+R),$$
or equivalently as
$$\beta(x,y)=-\zeta(x,y)-\int_x^\infty dz\,\zeta(x,z)\,
r(x;z,y),\tag 1.5$$
where $r(x;z,y)$ denotes the integral kernel of the operator $R.$

Let us consider (1.3) in the special case
$$\alpha+\omega+\alpha\Omega=0,\tag 1.6$$
where the nonhomogeneous term and the integral kernel coincide, as
seen by writing (1.6) explicitly as
$$\alpha(x,y)+\omega(x,y)+\int_x^\infty dz\,\alpha(x,z)\,\omega(z,y)=0,
\qquad y>x.\tag 1.7$$
The solution to (1.6) can formally be written as
$$\alpha=-\omega(I+R).\tag 1.8$$
The unique solvability of (1.7) in ${\Cal H}_1^{N\times N}$ and the condition
in (1.2) imply that
$$\sup_{y>x}\,\int_x^\infty dz\left(\|\alpha(z,y)\|+\|\alpha(y,z)\|\right)
<+\infty.\tag 1.9$$

A fundamental question related to (1.3) is the following: Can we
write $r(x;y,z)$ appearing in (1.5) explicitly in terms of $\alpha(x,y)$
appearing in (1.7)? In case the
answer is affirmative, we can express the solution
$\beta$ to (1.3) explicitly in terms of $\alpha$ and $\zeta.$
In fact, such a reduction question
dates back to the Armenian astrophysicist Ambarzumian whose invariance
principles are used in transfer of light in planetary atmospheres [7,11,12,32,33].
Ambarzumian [6] considered (1.6) with $\omega(y,z)=(c/2)\text{Ei}(|y-z|),$
where $\text{Ei}$ is the exponential integral function.
Similar reduction formulas were obtained [20,21] for integral
equations with convolution kernels, i.e. when
$\omega(y,z)$ is a function of $(y-z).$

One of our goals in this paper is to study the aforementioned fundamental question
when the integral operator $\Omega$ in (1.3) is $N\times N$-matrix valued
and $J$-selfadjoint in the sense that
$$\Omega=J\Omega^\dagger J,\quad \omega(y,z)=J\,\omega(z,y)^\dagger J,\tag 1.10$$
where the dagger denotes the matrix adjoint (complex conjugation and matrix
transpose) and $J$ is an $N\times N$ selfadjoint involution, i.e.
$$J=J^\dagger=J^{-1}.$$
We present one of our key results in Theorem 2.2, where the resolvent kernel $r(x;y,z)$
appearing in (1.5) is explicitly expressed in terms of the solution
$\alpha(x,y)$ to (1.6).

Let us note that, without loss of generality, $J$ may be assumed to have the form
$$J:=\bmatrix I_j&0\\
\noalign{\medskip}
0&-I_{N-j}\endbmatrix,$$
where $I_j$ is the $j\times j$ identity matrix for some $1\le j\le N.$
In that case we have
$$J\bmatrix M_1&M_2\\
\noalign{\medskip}
M_3&M_4\endbmatrix J=\bmatrix M_1&-M_2\\
\noalign{\medskip}
-M_3&M_4\endbmatrix,$$
for block matrices $M_1,$ $M_2,$ $M_3,$ $M_4$ of appropriate sizes.

Having established our first key result in Theorem 2.2, we turn our
attention to the integral equation
$$\tilde\alpha+\tilde\omega+\tilde\alpha\tilde\Omega=0,\tag 1.11$$
which is explicitly written as
$$\tilde\alpha(x,y)+\tilde\omega(x,y)+\int_x^\infty dz\,\tilde\alpha(x,z)\,
\tilde\omega(z,y)=0,\qquad y>x,\tag 1.12$$
obtained from (1.6) by perturbing the operator $\Omega$ to
$\tilde\Omega$
by a finite-rank operator, i.e.
$$\tilde\Omega=\Omega+FG,\quad
\tilde\omega(x,y)=\omega(x,y)+f(x)\,g(y),\tag 1.13$$
where $f$ and $g$ are $N\times j$ and $j\times N$ matrices with entries depending
on a single independent variable and belonging to
${\Cal H}_1^{N\times j}\cap{\Cal H}_\infty^{N\times j}$ and
${\Cal H}_1^{j\times N}\cap{\Cal H}_\infty^{j\times N}$, respectively.
We note that we cannot in general expect $F$ and $G$ to commute, and hence
in general $fg\ne gf.$
In our second key result, we show that (1.11) can be transformed into another
integral equation
in which the kernel is degenerate (i.e. separable
in the independent variables) so that $\tilde\alpha(x,y)$ can be explicitly
obtained in terms of $\alpha(x,y),$ $f(x),$ and $g(y),$ as
indicated in Theorem 3.4.

Our key result given in Theorem 3.4
has important implications for
various linear differential equations or systems of differential equations
arising in important physical applications.
One important consequence of Theorem 3.4 is that it provides a systematic
method to derive the Darboux transformations for a wide variety of spectral
problems for differential equations. Recall that the idea behind
a Darboux transformation (see e.g. [9,27,31] and
the references therein) is to determine
how the (generalized) eigenvectors change when
a finite number of
discrete eigenvalues are
added to or subtracted from the spectrum of a differential operator without
changing the continuous spectrum. In the language of physics, the
Darboux transformation provides the perturbed potential
and wave function in terms of the
unperturbed quantities when a finite
number of bound states are added or subtracted.

In this paper we are only concerned with Darboux transformations and not with
B\"acklund transformations. When a discrete eigenvalue is added to the spectrum
of a differential operator,
a B\"acklund transformation [18] usually consists of a first-order
differential equation (or a system of first-order differential equations)
involving the perturbed and unperturbed potentials.
On the other hand, in a Darboux transformation the perturbed
potential is explicitly expressed in terms of unperturbed quantities.
B\"acklund transformations have been derived [8,13,19,25,34] for various
systems of differential operators and they are useful
in obtaining exact solutions to related nonlinear evolution equations
such as the Korteweg-de Vries equation,
the nonlinear Schr\"odinger equation, the sine-Gordon equation, and
the modified Korteweg-de Vries equation.

The Darboux transformation is well
understood for Sturm-Liouville problems on a finite interval [14] and the
one-dimensional Schr\"odinger equation [15], but there are also
many others where such transformations are not yet known or only
some very special cases are known.
Our Theorem 3.4 can be applied
to the Zakharov-Shabat differential operator, matrix Zakharov-Shabat systems, and
other differential operators to derive in a systematic way the corresponding
Darboux transformations both at the potential and wave function levels.
Our theorem is general enough so that it applies when one eigenvalue is added or
subtracted from the spectrum, several eigenvalues are added or subtracted
simultaneously, and eigenvalues with nontrivial Jordan structures are added
or subtracted either one at a time or simultaneously. As an example,
we apply Theorem 3.4 on the Zakharov-Shabat system, and in Theorem 6.4 we
present the Darboux transformation expressing
both the change in the potential and the change in the wave function
explicitly in terms of the wave function of the unperturbed problem
when one bound state is added.
We compare our transformation for the potential given in (6.18)
and for the wave function given in (6.19)
with the results in the literature.

Let us mention that our results remain valid if the range of the integral is
over $(-\infty,x)$ so that (1.7) is replaced with
$$\alpha(x,y)+\omega(x,y)+\int_{-\infty}^x dz\,\alpha(x,z)\,\omega(z,y)=0,
\qquad y<x,\tag 1.14$$
and also valid if the integral is over $(0,x)$ so that (1.7) is replaced with
$$\alpha(x,y)+\omega(x,y)+\int_0^x dz\,\alpha(x,z)\,\omega(z,y)=0,
\qquad 0<y<x,\tag 1.15$$
and with the obvious appropriate replacements in (1.1),
(1.2), (1.5), (1.7), (1.9), and (1.12).
By using the operator notation of (1.3) and (1.6) it is straightforward
to modify the proofs and to
treat (1.7), (1.14), and (1.15) all at once.

Our paper is organized as follows. In Section 2 we establish our first
key result by expressing the resolvent kernel $r(x;y,z)$ appearing in
(1.5) explicitly in terms of the solution $\alpha(x,y)$ to (1.6).
In Section 3 we obtain
our second key result by expressing $\tilde\alpha(x,y)$ explicitly in terms of
$\alpha(x,y),$ $f(x),$ $g(y)$ when (1.6) is perturbed to (1.11) as in (1.13).
In Section 4 we show how the key result in Theorem 3.4 provides
a unified approach to derive Darboux transformations.
In Section 5 we show that the key results in the previous
sections are applicable to various systems such as the
Zakharov-Shabat system, its matrix generalizations, and the
Schr\"odinger equations on the full and half lines;
we show how an integral equation
of the form (1.7), (1.14), or (1.15) arises
for each system and is related to
an associated Marchenko integral equation or a Gel'fand-Levitan integral
equation.
In Section 6, we illustrate the significance of
our Theorem 3.4 and derive the Darboux transformation for the
Zakharov-Shabat system and make a comparison
with some related results in the literature.

\vskip 10 pt
\noindent {\bf 2. REDUCTION OF THE RESOLVENT KERNEL}
\vskip 3 pt

Recall that we assume that (1.3) is uniquely solvable in ${\Cal H}_1^{N\times N}$
and in ${\Cal H}_2^{N\times N}$ and that the operator $\Omega$ and its integral kernel
$\omega(y,z)$ satisfy (1.2) and (1.10).
In this section, we analyze the resolvent kernel $r(x;y,z)$ appearing in (1.5) and
present our first key
result; namely, we show that
$r(x;y,z)$ can be expressed explicitly in terms of the solution $\alpha(x,y)$ to (1.7).

\noindent {\bf Proposition 2.1} {\it Assume that (1.3) is uniquely
solvable in ${\Cal H}_2^{N\times N}$ and that $\Omega$ satisfies (1.10). Then,
the operator $R$ given in (1.4) and the corresponding
kernel $r(x;y,z)$ appearing in (1.5) satisfy
$$R=J R^\dagger J,\quad r(x;y,z)=J\,r(x;z,y)^\dagger J,\tag 2.1$$
where $J$ is the involution matrix appearing in (1.10).}

\noindent PROOF: From (1.4) we see that
$$(I+\Omega)(I+R)=I=(I+R)(I+\Omega),$$
and hence we obtain
$$R+\Omega+ \Omega R=0,\tag 2.2$$
$$R+\Omega+ R\Omega=0.\tag 2.3$$
By taking the adjoint of the operator equation in (2.2) and
applying $J$ on both sides of the resulting equation, we get
$$J R^\dagger J+J\Omega^\dagger J+ (J R^\dagger J)(J \Omega^\dagger J)=0,$$
or equivalently, after using (1.10),
$$J R^\dagger J+\Omega+ (J R^\dagger J)\Omega=0.\tag 2.4$$
Since (1.3) is assumed to be uniquely solvable in ${\Cal H}_2^{N\times N}$,
by comparing (2.3) and (2.4) we see that $R=JR^\dagger J.$
In taking the adjoint, we note that the independent variables $y$ and $z$ are switched
in the argument of the kernel and hence (2.1) is established. \qed

Our first key result is given in the next theorem.

\noindent {\bf Theorem 2.2} {\it
Assume that (1.3) is uniquely
solvable in ${\Cal H}_2^{N\times N}$ and that $\Omega$ satisfies (1.10). Then,
the corresponding
kernel $r(x;y,z)$ appearing in (1.5) can be expressed
explicitly in terms of the solution $\alpha(x,y)$ to (1.7) as}
$$r(x;y,z)=\cases
\alpha(y,z)+\displaystyle\int_x^y
ds\,J\,\alpha(s,y)^\dagger J\,\alpha(s,z),\qquad x<y<z,\\
\noalign{\medskip}
J\,\alpha(z,y)^\dagger J
+\displaystyle\int_x^z
ds\,J\,\alpha(s,y)^\dagger J\,\alpha(s,z),\qquad x<z<y,\endcases\tag 2.5$$
{\it where $J$ is the involution matrix appearing in (1.10).}

\noindent PROOF: Since (1.3) is uniquely solvable in ${\Cal H}_2^{N\times N}$,
so is (1.6) and hence the solution $R$ to (2.3) is unique. Thus, it suffices
to prove that the quantity defined in (2.5) satisfies (2.3), i.e. the quantity
in (2.5) satisfies the integral equation
$$r(x;y,z)+\omega(y,z)+\int_x^\infty ds\,
r(x;y,s)\,\omega(s,z)=0,\qquad x<\text{min}\{y,z\}.\tag 2.6$$
The proof for the case $x<z<y$ is similar to the case $x<y<z,$ and hence
we will only give the proof in the latter case. In that case,
let us use $\int_x^\infty=\int_x^y+\int_y^\infty$
in the integral appearing in (2.6).
We use a direct substitution from (2.5) into (2.6),
where we note that the first line of (2.5) is used in
the integral $\int_y^\infty$ and the second line
of (2.7) is used in the
integral $\int_x^y$ in (2.6).
After the substitution,
the left hand side in (2.6)
becomes
$v_1+v_2+v_3,$
where we have defined
$$v_1:=\alpha(y,z)+\omega(y,z)+
\int_y^\infty ds\,\alpha(y,s)\,\omega(s,z),$$
$$v_2:=\displaystyle\int_x^y
dt\,
J\,\alpha(t,y)^\dagger J\,
\omega(t,z)+
\displaystyle\int_x^y
dt\,J\,\alpha(t,y)^\dagger J\,\alpha(t,z),$$
$$v_3:=\displaystyle\int_x^y ds \int_x^s dt\,
J\,\alpha(t,y)^\dagger J\,\alpha(t,s)\,\omega(s,z)+
\int_y^\infty ds \displaystyle\int_x^y dt\,
J\,\alpha(t,y)^\dagger J\,\alpha(t,s)\,\omega(s,z).$$
Note that $v_1=0$ because of (1.7).
The orders of the two iterated integrals in $v_3$ can be
changed to
$\int_x^y dt\int_t^y ds$ and
$\int_x^y dt\int_y^\infty ds,$ respectively. Using
$\int_t^y +\int_y^\infty=\int_t^\infty,$ we then get
$$v_2+v_3=\displaystyle\int_x^y dt\,
J\,\alpha(t,y)^\dagger J
\left[\alpha(t,z)+\omega(t,z)+\int_t^\infty ds\,\alpha(t,s)\,\omega(s,z)
\right].\tag 2.7$$
We see that the quantity in the brackets in (2.7) vanishes because
of (1.7). Thus, (2.6) is satisfied for $x<y<z.$ A similar direct
substitution for the case $x<z<y$ completes the proof. \qed

\vskip 10 pt
\noindent {\bf 3. FINITE-RANK PERTURBATIONS}
\vskip 3 pt

Our main goal in this section is to show that the solution
$\tilde\alpha(x,y)$ to (1.12) can be expressed explicitly
in terms of $\alpha(x,y),$ $f(x),$ and $g(y)$ appearing in
(1.7) and (1.13), respectively. As we will see in later
sections, the key formulas given in (3.19)-(3.21) below
form the basis of Darboux transformations related to a wide variety of
spectral problems associated with ordinary differential operators.

Recall that we assume that
(1.10) holds and that
(1.7) is uniquely solvable on
${\Cal H}_1^{N\times N}$ and on ${\Cal H}_2^{N\times N}.$
Let us now define the intermediate quantities $n(x)$ and
$q(y)$ as
$$n(x):=f(x)+\int_x^\infty dz\,\alpha(x,z)\,f(z),\quad
q(x):=g(x)+\int_x^\infty dz\,g(z)\,J\,\alpha(x,z)^\dagger\,J,\tag 3.1$$
where $J$ is the involution matrix appearing in (1.10). From (1.9) it follows that
$n\in{\Cal H}_1^{N\times j}\cap{\Cal H}_\infty^{N\times j}$ and
$q\in{\Cal H}_1^{j\times N}\cap{\Cal H}_\infty^{j\times N}.$
Note that both integration limits $\int_x^\infty$ in (3.1) can be replaced with
$\int_{-\infty}^\infty$ because $\alpha(x,y)=0$ for $x>y.$

\noindent {\bf Theorem 3.1} {\it We can transform (1.11) into an
integral equation that has a degenerate kernel and hence obtain
$\tilde\alpha$ explicitly by linear algebraic methods.}

\noindent PROOF: Using (1.13) let us write (1.11) as
$$\tilde\alpha(I+\Omega+FG)=-\omega-fg.\tag 3.2$$
Recall that all the operators act from the right. By applying on (3.2) from the
right with the resolvent operator
$(I+R)$ appearing in (1.4), we get
$$\tilde\alpha[I+FG(I+R)]=\alpha-fg(I+R),\tag 3.3$$
where we have used (1.8) to have $\alpha$ on the right hand side of (3.3).
Let us define the operator $\tilde G$ as
$$\tilde G:=G(I+R),\quad \tilde g(x,y):=g(y)+\int_x^\infty dz\,g(z)\,r(x;z,y),
\tag 3.4$$
where $r(x;y,z)$ is the kernel given in (2.5).
We emphasize the dependence of $\tilde g$ both on $x$ and $y.$
Note that the integral equation in (3.3) has a degenerate kernel, which can be
seen by writing it in the form
$$\tilde\alpha(I+F\tilde G)=\alpha-f\tilde g,\tag 3.5$$
because the kernel of $F\tilde G$ is $f(y)\,\tilde g(x,z),$ where there
is a separation of the $y$ and $z$ variables, and $x$
appears merely as a parameter.

Let us now solve (3.3) by using
linear algebra. We look for a solution in the form
$$\tilde\alpha(x,y)=\alpha(x,y)+p(x)\,\tilde g(x,y),\tag 3.6$$
where $p$ is to be determined.
Using (3.6) in (3.5), after some simplification we get
$$(\alpha F+p+p \tilde g F+f)\tilde G=0,$$
which yields
$$p(I+\tilde g F)=-(f+\alpha F),\tag 3.7$$
or written in the integral form as
$$p(x)=-n(x)\left[I+\int_x^\infty
ds\,\tilde g(x,s)\,f(s)\right]^{-1},$$
where we have used the definition of $n(x)$ given in (3.1).
Using (3.7) in (3.6) we obtain
$$\tilde\alpha=\alpha-(f+\alpha F)(I+\tilde g F)^{-1}\tilde g,$$
or written in the integral form as
$$\tilde\alpha(x,y)=\alpha(x,y)-
n(x)\left[I+\int_x^\infty
ds\,\tilde g(x,s)\,f(s)\right]^{-1}
\tilde g(x,y),\tag 3.8$$
which completes the proof of our theorem. \qed

Note that (3.8) expresses $\tilde\alpha(x,y)$ in terms of
$\alpha(x,y),$ $f(x),$ and $\tilde g(x,y)$
because as seen from (3.1) the quantity $n(x)$
is available in terms of $\alpha(x,y)$ and $f(x).$

Next, we show that $\tilde g(x,y)$ can explicitly
be obtained in terms of
$\alpha(x,y)$ and $g(y),$ which will then imply that
$\tilde\alpha(x,y)$ is expressed in terms of
$\alpha(x,y),$ $f(x),$ and $g(y).$

\noindent {\bf Proposition 3.2} {\it The quantity
$\tilde g(x,y)$ defined in (3.4) can be expressed
explicitly in terms of the solution $\alpha(x,y)$ to (1.6) and
the quantities $f(x)$ and $g(y)$ appearing in (1.13) as}
$$\tilde g(x,y)=q(y)+\int_x^y ds\,q(s)\,\alpha(s,y),\tag 3.9$$
{\it where $q(y)$ is the quantity defined in (3.1),
and it is noted that $\tilde g(x,x)=q(x)$.}

\noindent PROOF: We will substitute (2.5) into (3.4). For this purpose,
let us write the integral $\int_x^\infty$ in (3.4) as $\int_x^y+\int_y^\infty.$
Using the first line of (2.5) in $\int_x^y$ and the second in $\int_y^\infty,$
we obtain
$$\aligned
\tilde g(x,y)=&g(y)+\int_x^y ds\,g(s)\,\alpha(s,y)+\int_y^\infty ds\,
g(s)\,J\,\alpha(y,s)^\dagger\,J\\
&+\left(\int_x^y ds\int_x^s dt+ \int_y^\infty ds\int_x^y dt \right)
g(s)\,J\,\alpha(t,s)^\dagger\,J\,\alpha(t,y).\endaligned$$
The sum of the two iterated integrals above can be written first as a double integral
and then as an iterated integral by changing the order of integration to get
$$\int_x^y ds\int_x^s dt+ \int_y^\infty ds\int_x^y dt
=\int_x^y dt\int_t^\infty ds.\tag 3.10$$
Using (3.10) and combining terms as in (3.1), we then obtain (3.9). \qed

We note that the integral $\int_x^y$ in (3.9) can also be written as
$\int_x^\infty$ because $\alpha(s,y)=0$ for $s>y.$

Let us define the matrix $\Gamma(x)$ as the quantity whose inverse appearing
in (3.8), namely as
$$\Gamma(x):=I+\int_x^\infty ds\,\tilde g(x,s)\,f(s).\tag 3.11$$

\noindent {\bf Proposition 3.3} {\it The quantity
$\Gamma(x)$ defined in (3.11) can be expressed
explicitly in terms of the solution $\alpha(x,y)$ to (1.6) and
the quantities $f(x)$ and $g(y)$ appearing in (1.13) as}
$$\Gamma(x)=I+\int_x^\infty ds\,q(s)\,n(s),\tag 3.12$$
{\it where $n(x)$ and $q(x)$ are the quantities defined in (3.1).}

\noindent PROOF: Using (3.9) in (3.11) we get
$$\Gamma(x)=I+\int_x^\infty ds\,q(s)\,f(s)
+\int_x^\infty ds\,\int_x^s dt\, q(t)\,\alpha(t,s)\,f(s).\tag 3.13$$
Changing the order of integration in the last integral in (3.13)
and using (3.11), we get (3.12). Since $n(s)$ and $q(s)$ are
expressed in terms of $\alpha(x,y),$
$f(x),$ and $g(y),$ we see from (3.12) that
$\Gamma(x)$ is explicitly expressed in terms of $\alpha(x,y),$
$f(x),$ and $g(y),$ as well. \qed

The Fourier transform
of the $N\times N$ matrix-valued quantity $\alpha(x,y)$ in (1.7), usually called a
wave function, can be written as
$$\Psi(\lambda,x):=e^{-i\lambda Jx}+\int_x^\infty dy\,\alpha(x,y)e^{-i\lambda Jy},
\tag 3.14$$
where $J$ is the involution matrix appearing in (1.10).
Using the inverse Fourier transform on (3.14) we get
$$\alpha(x,y)=\displaystyle\frac{1}{2\pi}\int_{-\infty}^\infty d\lambda\,
\left[\Psi(\lambda,x)-e^{-i\lambda Jx}\right]e^{i\lambda Jy}.\tag 3.15$$
Similarly, for $\tilde\alpha(x,y),$ we have the associated $N\times N$
matrix-valued wave function $\tilde\Psi(\lambda,x),$ where
$$\tilde\Psi(\lambda,x):=e^{-i\lambda Jx}+\int_x^\infty dy\,
\tilde\alpha(x,y)e^{-i\lambda Jy},\tag 3.16$$
$$\tilde\alpha(x,y)=\displaystyle\frac{1}{2\pi}\int_{-\infty}^\infty d\lambda\,
\left[\tilde\Psi(\lambda,x)-e^{-i\lambda Jx}\right]e^{i\lambda Jy}.$$
Let us introduce
$$\gamma(\lambda,x):=\int_x^\infty dy\,\tilde g(x,y)e^{-i\lambda Jy}.\tag 3.17$$
Using (3.9) and (3.14) in (3.17)
and the fact that $\alpha(x,y)=0$ for $y<x,$ we get
$$\gamma(\lambda,x)=\int_x^\infty ds\,q(s)\Psi(\lambda,s).\tag 3.18$$

The following theorem and in particular (3.20) describes the effect of
a finite-rank perturbation on the wave function.

\noindent {\bf Theorem 3.4} {\it Let $\alpha$ and
$\tilde\alpha$ be the solutions to the integral equations (1.6) and
(1.11), respectively, and let $n(x),$ $\Gamma(x),$ \ $\tilde g(x,y),$
and $\gamma(\lambda,x)$
be the quantities given in (3.1), (3.12), (3.9), and (3.18), respectively. Then,
$\tilde\alpha(x,y)-\alpha(x,y)$
and $\tilde{\Psi}(\lambda,x)-\Psi(\lambda,x)$ can explicitly
be written in terms of $\alpha(x,y),$ $f(x),$ $g(y)$ as}
$$\tilde \alpha(x,y)-\alpha(x,y)=-n(x)\,\Gamma(x)^{-1}\tilde g(x,y),\tag 3.19$$
$$\tilde{\Psi}(\lambda,x)-\Psi(\lambda,x)=-n(x)\Gamma(x)^{-1}\gamma(\lambda,x).
\tag 3.20$$
{\it Furthermore, we have}
$$\tilde \alpha(x,x)-\alpha(x,x)=-n(x)\,\Gamma(x)^{-1}q(x).\tag 3.21$$

\noindent PROOF:
Note that using (3.1), (3.11), and (3.12) in (3.8) we obtain
(3.19). Using $\tilde g(x,x)=q(x)$ from Proposition~3.2 in (3.19) we get (3.21).
Finally, we obtain (3.20)
with the help of (3.14), (3.16), (3.17), and (3.19). \qed
\medskip

We conclude this section with
a result on the trace of the left hand side of
(3.21).

\noindent {\bf Proposition 3.5} {\it Let $\alpha$ and
$\tilde\alpha$ be the solutions of the integral equations (1.6) and
(1.11), respectively, and let $\Gamma$ be the matrix
given in (3.12).
The trace of the difference
$\tilde \alpha(x,x)-\alpha(x,x)$ can be expressed as
the logarithmic derivative of the determinant of $\Gamma(x)$ as}
$$\text{tr}\left[\tilde \alpha(x,x)-\alpha(x,x)\right]=\text{tr}\left[
\displaystyle\frac{d\Gamma(x)}{dx}\,\Gamma(x)^{-1}\right]=
\displaystyle\frac{1}{\det \Gamma(x)}\displaystyle\frac{d\det \Gamma(x)}{dx}.\tag 3.22$$

\noindent PROOF: From (3.12) we get
$$\displaystyle\frac{d\Gamma(x)}{dx}=-q(x)\,n(x).\tag 3.23$$
Using the well-known matrix properties
$$\text{tr}\left[M_1 M_2
\right]=\text{tr}\left[M_2 M_1
\right],\quad \text{tr}\left[\displaystyle\frac{d M(x)}{dx}\,M(x)^{-1}\right]
=
\displaystyle\frac{1}{\det M(x)}\displaystyle\frac{d\det M(x)}{dx},$$
 from (3.21) and (3.23) we get (3.22). \qed

\vskip 10 pt
\noindent {\bf 4. DARBOUX TRANSFORMATIONS}
\vskip 3 pt

The integral equations (1.7), (1.14), and (1.15)
arise in the study of various scattering
and spectral problems, some of which
are described in Section 5. In this section we will elaborate
on (3.19)-(3.21) and show how
they provide a unified approach
to Darboux transformations for a variety of scattering and spectral problems.
Recall that a Darboux transformation describes how the wave function and
the potential change when a finite number of
discrete eigenvalues are added (or subtracted) from the spectrum of a
differential operator without changing its continuous spectrum.

Suppose we add a discrete eigenvalue $\lambda_j$ with multiplicity $n_j$
to the existing spectrum. Then, associated with the eigenvalue
$\lambda_j,$ there are $n_j$ parameters $c_{j0},\dots,c_{j(n_j-1)},$
usually known as norming constants.
The formulas (3.19)-(3.21) tell us how the wave function
changes from $\Psi(\lambda,x)$ to $\tilde \Psi(\lambda,x),$
how the potential changes from $u(x)$ to $\tilde u(x),$ and
how the quantity $\alpha(x,y)$ related to the Fourier transform of
the wave function changes to $\tilde \alpha(x,y).$
Consequently, for each discrete eigenvalue $\lambda_j$ added to the spectrum,
there will be an $n_j$-parameter family of potentials
$\tilde u(x),$ where the norming constants act as the parameters.
In case several discrete eigenvalues $\lambda_1,\dots,\lambda_N$
are added all at once, it is convenient to use [4]
a square matrix $A$ whose eigenvalues are related to $\lambda_j$ for $j=1,\dots,N$
in a simple manner; it is also convenient to use [4]
a matrix $C$ whose entries are related to the norming constants $c_{js}$
for $j=1,\dots,N$ and $s=0,1,\dots,n_j-1.$

The quantities $f(x)$ and $g(x)$ appearing in (1.13) can usually be
represented in the form
$$f(x)=\bmatrix 0&B^\dagger e^{-A^\dagger x}\\
\noalign{\medskip}
Ce^{-Ax}&0\endbmatrix,\quad
g(y)=\bmatrix e^{-Ay}B&0\\
\noalign{\medskip}
0&-e^{-A^\dagger y}C^\dagger\endbmatrix,\tag 4.1$$
where $A$ is a constant square matrix with
all eigenvalues having positive real parts (the bound-state $\lambda$-values
are usually obtained [4] by multiplying the eigenvalues of $A$ by the
imaginary unit $i$), and
$B$ and $C$ are constant matrices of appropriate
sizes so that the matrix product $f(x)\,g(y)$ is well defined and given by
$$f(x)\,g(y)=\bmatrix 0&-B^\dagger e^{-A^\dagger (x+y)}
C^\dagger\\
\noalign{\medskip}
Ce^{-A(x+y)}B&0\endbmatrix.$$

For $f(x)$ and $g(y)$ given in (4.1),
let us evaluate $n(x),$ $q(x),$ and $\tilde g(x,y)$ given in (3.1) and
(3.9), respectively, explicitly in terms of
the wave function $\Psi(\lambda,x)$ evaluated at the eigenvalues of
$A.$
First, by taking the matrix adjoint,
 from (3.15) we get
$$J\alpha(x,y)^\dagger
J=\displaystyle\frac{1}{2\pi}\int_{-\infty}^\infty d\lambda\,e^{-i\lambda Jy}
\left[J\Psi(\lambda,x)^\dagger J-e^{i\lambda Jx}\right].\tag 4.2$$
Using (3.15) in (3.1) and the fact that
$$\displaystyle\frac{1}{2\pi}\int_{-\infty}^\infty ds\,e^{\pm i a s}=\delta(a),
\tag 4.3$$
where $\delta$ is the Dirac delta distribution, we
evaluate $n(x)$ as
$$n(x)=\displaystyle\frac{1}{2\pi}\int_{-\infty}^\infty d\lambda\,
\Psi(\lambda,x)\int_x^\infty dy
\,e^{i\lambda Jy}f(y).\tag 4.4$$
Using (4.1) in (4.4) we obtain
$$n(x)=\displaystyle\frac{1}{2\pi i}\int_{-\infty}^\infty d\lambda\,
\Psi(\lambda,x)\,\,e^{i\lambda Jx}\Cal N(\lambda,x),\tag 4.5$$
where we have defined
$$\Cal N(\lambda,x):=
\bmatrix 0&-B^\dagger (\lambda I+i A^\dagger)^{-1}
e^{-A^\dagger x}
\\
\noalign{\medskip}
C (\lambda I-i A)^{-1}e^{-A x} &0\endbmatrix.$$
Similarly, using (4.2) and (4.3) in (3.1) we obtain
$$q(x)=
\displaystyle\frac{1}{2\pi}\int_{-\infty}^\infty d\lambda\int_x^\infty
dy\,g(y)\,e^{-i\lambda Jy}
J\Psi(\lambda,x)^\dagger J,\tag 4.6$$
and using (4.1) in (4.6) we conclude that
$$q(x)=
\displaystyle\frac{1}{2\pi i}\int_{-\infty}^\infty
d\lambda\,
\Cal Q(\lambda,x)e^{-i\lambda J x}J\Psi(\lambda,x)^\dagger J,\tag 4.7$$
where we have defined
$$\Cal Q(\lambda,x):=
\bmatrix e^{-A x}(\lambda I-i A)^{-1}B
& 0\\
\noalign{\medskip}
0&e^{-A^\dagger x}(\lambda I+i A^\dagger)^{-1}
C^\dagger
\endbmatrix.$$

Proceeding in a similar manner, with the help of
(3.9), (3.14), (4.3), and (4.7) we first get
$$\tilde g(x,y)=\displaystyle\frac{1}{2\pi}\int_{-\infty}^\infty d\mu
\int_x^\infty dz\,q(z)\,\Psi(\mu,z)\,e^{i\mu Jy},$$
and then obtain
$$\tilde g(x,y)=\int_{-\infty}^\infty d\lambda
\int_{-\infty}^\infty d\mu
\int_x^\infty dz\,E(\lambda,\mu,z)\,e^{i\mu Jy},\tag 4.8$$
where we have defined
$$E(\lambda,\mu,x):=\displaystyle\frac{1}{4\pi^2 i}\,\Cal Q(\lambda,x)\,e^{-i\lambda Jx}
J\Psi(\lambda,x)^\dagger J\,
\Psi(\mu,x).\tag 4.9$$
Then, with the help of (4.5) and (4.7)
the matrix $\Gamma(x)$ given in (3.12) can be explicitly written
in terms of the wave function $\Psi$ as
$$\Gamma(x)=I-i
\int_{-\infty}^\infty d\lambda
\int_{-\infty}^\infty d\mu\int_x^\infty dy\,E(\lambda,\mu,y)\,\,
e^{i\mu Jy}\Cal N(\mu,y).
\tag 4.10$$
Finally, using (4.7) and (4.9) in (3.18), we obtain
$$\gamma(\lambda,x)=2\pi\int_x^\infty ds\int_{-\infty}^\infty d\lambda\,
E(\lambda,\lambda,s).\tag 4.11$$

Note that the integrals in (4.5), (4.7), (4.10), and (4.11) can be performed
as residue integrals in the complex
$\lambda$-plane with the poles at the eigenvalues of
$iA$ and $-iA^\dagger.$
Evaluating those integrals and using the result in (4.8) and (4.11)
we can obtain $\tilde g(x,y)$ and $\gamma(\lambda,x)$
explicitly in terms of $\Psi(\lambda,x)$ evaluated at the
eigenvalues of $iA$ and $-iA^\dagger.$
If some bound states have multiplicities, i.e. if
some of the eigenvalues of $A$ have nontrivial Jordan structures,
then the explicit expressions for $\tilde g(x,y)$ and $\gamma(\lambda,x)$ also contain
some $\lambda$-derivatives of $\Psi(\lambda,x)$ evaluated at the
eigenvalues of $iA$ and $-iA^\dagger.$

In Section 6 we will use the procedure described here to obtain the Darboux transformation
for the Zakharov-Shabat system given in (5.1).

\vskip 10 pt
\noindent {\bf 5. APPLICATIONS TO SPECIFIC SYSTEMS}
\vskip 3 pt

In this section we present some specific systems on which the theory presented
in the previous sections is applicable. In the first four examples
we analyze the Zakharov-Shabat system and its matrix generalizations.
In the remaining three examples we analyze the Schr\"odinger equation on the full and
half lines.
For each system, we identify the quantities
$\omega(x,y)$ and $\alpha(x,y)$ appearing in
(1.7) or in one of its variants
(1.14) and (1.15), and we identify $f(x)$ and $g(y)$ appearing in (1.13).
We identify the involution $J$ for which (1.10) is satisfied by $\omega(y,z)$ for each system.
Then, for each system we indicate how
the potential $u(x)$ appearing in the system is related to $\alpha(x,x)$ so that
one sees clearly how the perturbation
$\tilde u(x)-u(x)$ in the potential can be recovered from (3.21).
For each system we indicate how the wave function $\Psi(\lambda,x)$
is related to certain specific solutions to
the corresponding system. We also relate the integral equations (1.7), (1.14), and (1.15)
to the Marchenko equations or the Gel'fand-Levitan equation
corresponding to each system. In some cases
we observe that (1.7), (1.14), or (1.15) is exactly the same as
a Marchenko equation or a Gel'fand-Levitan equation, and in some other cases
one needs to rearrange the Marchenko equations or the Gel'fand-Levitan equation
in order to get the integral equation (1.7), (1.14), or (1.15).

We first discuss four examples involving the Zakharov-Shabat system and its
matrix generalization, two with the range of the integral over $(x,+\infty)$
and two with the range of the integral over $(-\infty,x)$. We shall discuss
the details of Example 5.1 in the next section, where we
derive the Darboux transformation for the
Zakharov-Shabat system and compare our results
with the existing results in the literature [10,19,23,24,26,29].

\noindent {\bf Example 5.1} Consider the Zakharov-Shabat system
$$\displaystyle\frac{d \varphi(\lambda,x)}{dx}=\bmatrix -i\lambda&u(x)\\
\noalign{\medskip}
-u(x)^*&i\lambda\endbmatrix \varphi(\lambda,x),\tag 5.1$$
where the asterisk denotes complex conjugation, $u$ is a (scalar)
complex-valued integrable potential, and $\varphi$ is a column vector with
two components. The corresponding (left) Marchenko equations are given by [30]
$$\cases\bar K(x,y)+\bmatrix 0\\
\noalign{\medskip}
\Omega_{\text l}(x+y)\endbmatrix+\displaystyle\int_x^\infty dz\,K(x,z)\,
\Omega_{\text l}(z+y)=\bmatrix 0\\
\noalign{\medskip}
0\endbmatrix,\qquad y>x,\\
\noalign{\medskip}
K(x,y)-\bmatrix \Omega_{\text l}(x+y)^\dagger\\
\noalign{\medskip}
0\endbmatrix-\displaystyle\int_x^\infty dz\,\bar K(x,z)\,\Omega_{\text l}(z+y)^\dagger=\bmatrix 0\\
\noalign{\medskip}
0\endbmatrix,\qquad y>x,\endcases\tag 5.2$$
where $K(x,y)$ and $\bar K(x,y)$ are $2\times 1$ matrix
valued and $\Omega_{\text l}(x+y)$ is a scalar function.
Let us stress that an overline does not
indicate complex conjugation.
The potential
is recovered as [30]
$$u(x)=-2\bmatrix 1&0\endbmatrix K(x,x)=2\,\bar K(x,x)^\dagger \bmatrix 0\\
\noalign{\medskip}
1\endbmatrix.\tag 5.3$$
By letting
$$\alpha(x,y)=\bmatrix \bar K(x,y)&K(x,y)\endbmatrix,\quad \omega(x,y)=\bmatrix
0&-\Omega_{\text l}(x+y)^\dagger\\
\noalign{\medskip}
\Omega_{\text l}(x+y)&0\endbmatrix,\tag 5.4$$
we can write (5.2) as (1.7), which is now a $2\times 2$
system of integral equations. We note that
$\omega(x,y)$ given in (5.4) satisfies (1.10) with
$J=\text{diag}[1,-1].$
As seen from (5.3), $u$ is then recovered from the solution to (1.7) as
$$u(x)=-2\bmatrix 1&0\endbmatrix \alpha(x,x)
\bmatrix 0\\
\noalign{\medskip}
1\endbmatrix.\tag 5.5$$
In this case, the degenerate
perturbation on $\Omega_{\text l}(x+y)$
is given [4] by $C_{\text l}e^{-A_{\text l}(x+y)}B_{\text l},$
where $A_{\text l}$ is a constant $p\times p$ matrix
with all eigenvalues having positive real parts,
$B_{\text l}$ is a constant $p\times 1$ matrix, and
$C_{\text l}$ is a constant $1\times p$ matrix.
The functions $f$ and $g$ appearing in (1.13) are then $2\times 2p$ and
$2p\times 2$ matrices, respectively,
given by
$$f(x)=\bmatrix 0&B_{\text l}^\dagger e^{-A_{\text l}^\dagger x}\\
\noalign{\medskip}
C_{\text l}e^{-A_{\text l}x}&0\endbmatrix,\quad
g(y)=\bmatrix e^{-A_{\text l}y}B_{\text l}&0\\
\noalign{\medskip}
0&-e^{-A_{\text l}^\dagger y}C_{\text l}^\dagger\endbmatrix
.\tag 5.6$$
For the Zakharov-Shabat system the wave function appearing in
(3.14) is the $2\times 2$ matrix given by
$$\Psi(\lambda,x)=\bmatrix \bar\psi_1(\lambda,x)&\psi_1(\lambda,x)\\
\noalign{\medskip}
\bar\psi_2(\lambda,x)&\psi_2(\lambda,x)\endbmatrix=\bmatrix \psi_2(\lambda^*,x)^*&\psi_1(\lambda,x)\\
\noalign{\medskip}
-\psi_1(\lambda^*,x)^*&\psi_2(\lambda,x)\endbmatrix,\tag 5.7$$
where
$\bmatrix \psi_1(\lambda,x)\\
\psi_2(\lambda,x)\endbmatrix$ is the Jost solution to (5.1) with the
asymptotics
$\bmatrix 0\\
e^{i\lambda x}\endbmatrix$ as $x\to+\infty$
and
$\bmatrix \bar\psi_1(\lambda,x)\\
\bar\psi_2(\lambda,x)\endbmatrix$ is the solution
behaving as
$\bmatrix e^{-i\lambda x}\\
0\endbmatrix+o(1)$ as $x\to+\infty.$

\noindent {\bf Example 5.2} Consider the matrix generalization
of Example 5.1 with the Zakharov-Shabat system
$$\displaystyle\frac{d \varphi(\lambda,x)}{dx}=\bmatrix -i\lambda I_m&u(x)\\
\noalign{\medskip}
-u(x)^\dagger&i\lambda I_n\endbmatrix \varphi(\lambda,x),\tag 5.8$$
where $u$ is an $m\times n$ matrix with integrable entries, and
$I_j$ is the $j\times j$ identity matrix.
The corresponding (left) Marchenko equations are given by
$$\cases\bar K(x,y)+\bmatrix 0_{mn}\\
\noalign{\medskip}
\Omega_{\text l}(x+y)\endbmatrix+\displaystyle\int_x^\infty dz\,K(x,z)\,
\Omega_{\text l}(z+y)=\bmatrix 0_{mn}\\
\noalign{\medskip}
0_{nn}\endbmatrix,\qquad y>x,\\
\noalign{\medskip}
K(x,y)-\bmatrix \Omega_{\text l}(x+y)^\dagger\\
\noalign{\medskip}
0_{nm}\endbmatrix-\displaystyle\int_x^\infty dz\,\bar K(x,z)\,
\Omega_{\text l}(z+y)^\dagger=\bmatrix 0_{mm}\\
\noalign{\medskip}
0_{nm}\endbmatrix,\qquad y>x,\endcases\tag 5.9$$
where $0_{jk}$
is the zero matrix of size $j\times k,$
and $K,$ $ \bar K,$ and $\Omega_{\text l}$ have
sizes $(m+n)\times n,$ $(m+n)\times m,$ and
$n\times m,$ respectively.
The $m\times n$ potential matrix $u$
is recovered from the solution to (5.9) as
$$u(x)=-2\bmatrix I_m&0_{mn}\endbmatrix K(x,x)=2\,\bar K(x,x)^\dagger \bmatrix 0_{mn}\\
\noalign{\medskip}
I_n\endbmatrix.\tag 5.10$$
By letting
$$\alpha(x,y)=\bmatrix \bar K(x,y)&K(x,y)\endbmatrix,\quad \omega(x,y)=\bmatrix
0_{mm}&-\Omega_{\text l}(x+y)^\dagger\\
\noalign{\medskip}
\Omega_{\text l}(x+y)&0_{nn}\endbmatrix,\tag 5.11$$
we can write (5.9) as (1.7), which is now an $(m+n)\times (m+n)$
system of integral equations. Note that $\omega(x,y)$ given in (5.11) satisfies
(1.10) with
$J=I_m\oplus(-I_n).$
As seen from (5.10), the potential $u$ is recovered from the solution to (1.7) as
$$u(x)=-2\bmatrix I_m&0_{mn}\endbmatrix \alpha(x,x)
\bmatrix 0_{mn}\\
\noalign{\medskip}
I_n\endbmatrix.$$
In this case, the degenerate
perturbation on the $n\times m$ matrix
quantity $\Omega_{\text l}(x+y)$
is given [4,16,17] by $C_{\text l}e^{-A_{\text l}(x+y)}B_{\text l},$
where $A_{\text l}$ is a constant $p\times p$ matrix
with all eigenvalues having positive real parts,
$B_{\text l}$ is a constant $p\times m$ matrix, and
$C_{\text l}$ is a constant $n\times p$ matrix.
The functions $f$ and $g$ appearing in (1.13) are then $(m+n)\times 2p$ and
$2p\times (m+n)$ matrices, respectively,
given by
$$f(x)=\bmatrix 0_{mp}&B_{\text l}^\dagger e^{-A_{\text l}^\dagger x}\\
\noalign{\medskip}
C_{\text l}e^{-A_{\text l}x}&0_{np}\endbmatrix,\quad
g(y)=\bmatrix e^{-A_{\text l}y}B_{\text l}&0_{pn}\\
\noalign{\medskip}
0_{pm}&-e^{-A_{\text l}^\dagger y}C_{\text l}^\dagger\endbmatrix
.$$
The wave function appearing in (3.14) has the form
$$\Psi(\lambda,x)=\bmatrix \bar\psi_1(\lambda,x)&\psi_1(\lambda,x)\\
\noalign{\medskip}
\bar\psi_2(\lambda,x)&\psi_2(\lambda,x)\endbmatrix,$$
where
$\bmatrix \psi_1(\lambda,x)\\
\psi_2(\lambda,x)\endbmatrix$ is the $(m+n)\times n$ Jost solution to (5.8) with the
asymptotics
$\bmatrix 0\\
e^{i\lambda x}I_n\endbmatrix$ as $x\to+\infty$
and
$\bmatrix \bar\psi_1(\lambda,x)\\
\bar\psi_2(\lambda,x)\endbmatrix$ is the $(m+n)\times m$ Jost solution to (5.8)
with the asymptotics
$\bmatrix
e^{-i\lambda x}I_m\\
0_{nm}\endbmatrix$ as $x\to+\infty.$

\noindent {\bf Example 5.3} For the Zakharov-Shabat system in
(5.1), the (right) Marchenko integral equations are given by
$$\cases\bar M(x,y)+\bmatrix \Omega_{\text r}(x+y)\\
\noalign{\medskip}
0\endbmatrix+\displaystyle\int_{-\infty}^x dy\,M(x,z)\,\Omega_{\text r}(z+y)=\bmatrix 0\\
\noalign{\medskip}
0\endbmatrix,\qquad y<x,\\
\noalign{\medskip}
M(x,y)-\bmatrix 0\\
\noalign{\medskip}
\Omega_{\text r}(x+y)^\dagger\endbmatrix
-\displaystyle\int_{-\infty}^x dy\,\bar M(x,z)\,\Omega_{\text r}(z+y)^\dagger=\bmatrix 0\\
\noalign{\medskip}
0\endbmatrix,\qquad y<x.\endcases\tag 5.12$$
The scalar potential $u$ is recovered from the solution to (5.12) as
$$u(x)=2\bmatrix 1&0\endbmatrix \bar M(x,x)=-2\,M(x,x)^\dagger \bmatrix 0\\
\noalign{\medskip}
1\endbmatrix.\tag 5.13$$
By letting
$$\alpha(x,y)=\bmatrix M(x,y)&\bar M(x,y)\endbmatrix,\quad \omega(x,y)=\bmatrix
0&\Omega_{\text r}(x+y)\\
\noalign{\medskip}
-\Omega_{\text r}(x+y)^\dagger&0\endbmatrix,\tag 5.14$$
we can transform (5.12) into (1.14),
which is a $2\times 2$ system of integral equations.
Note that $\omega(x,y)$ given in (5.14) satisfies (1.10)
with
$J=\text{diag}[1,-1].$
As seen from (5.13), $u$ is recovered from the solution to (1.14) as
$$u(x)=2\bmatrix 1&0\endbmatrix \alpha(x,x)
\bmatrix 0\\
\noalign{\medskip}
1\endbmatrix.$$
In this case,
$f(x)$ and $g(y)$ appearing in (1.13) are given by
$$f(x)=\bmatrix 0&C_{\text r}e^{A_{\text r}x}
\\
\noalign{\medskip}
B_{\text r}^\dagger e^{A_{\text r}^\dagger x}&0\endbmatrix,\quad
g(y)=\bmatrix -e^{A_{\text r}^\dagger y}C_{\text r}^\dagger&0\\
\noalign{\medskip}
0&e^{A_{\text r}y}B_{\text r}
\endbmatrix,$$
where
$A_{\text r}$ is a constant $p\times p$ matrix
with all eigenvalues having positive real parts,
$B_{\text r}$ is a constant $p\times 1$ matrix, and
$C_{\text r}$ is a constant $1\times p$ matrix.
In this case the wave function appearing in
(3.14) is given by
$$\Psi(\lambda,x)=\bmatrix \phi_1(\lambda,x)&\bar\phi_1(\lambda,x)\\
\noalign{\medskip}
\phi_2(\lambda,x)&\bar\phi_2(\lambda,x)\endbmatrix=
\bmatrix \phi_1(\lambda,x)&-\phi_2(\lambda^*,x)^*\\
\noalign{\medskip}
\phi_2(\lambda,x)&\phi_1(\lambda^*,x)^*\endbmatrix,$$
where
$\bmatrix \phi_1(\lambda,x)\\
\phi_2(\lambda,x)\endbmatrix$ is the Jost solution to (5.1) with the
asymptotics
$\bmatrix
e^{-i\lambda x}\\
0\endbmatrix$ as $x\to-\infty$
and $\bmatrix \bar\phi_1(\lambda,x)\\
\bar\phi_2(\lambda,x)\endbmatrix$ is the solution behaving as
$\bmatrix
0\\
e^{i\lambda x}\endbmatrix+o(1)$ as $x\to-\infty.$

\noindent {\bf Example 5.4} Example 5.3 is also valid for the Zakharov-Shabat
system given in (5.8),
where $M,$ $\bar M,$ $\Omega_{\text r},$ $A_{\text r},$ $B_{\text r},$
$C_{\text r},$ $f,$ $g$ have now sizes $(m+n)\times m,$
$(m+n)\times n,$
$m\times n,$ $p\times p,$ $p\times n,$ $m\times p,$ $(m+n)\times 2p,$
$2p\times (m+n),$ respectively, and the eigenvalues of
the constant matrix
$A_{\text r}$ have all positive real parts, and
$B_{\text r}$ and
$C_{\text r}$ are constant matrices.
The matrix system in (5.12) can be written
as in (1.14), which is now an $(m+n)\times (m+n)$ system of integral equations
The $m\times n$ potential matrix $u$ is recovered from the solution to (1.14) as
$$u(x)=2\bmatrix I_m&0_{mn}\endbmatrix \alpha(x,x)
\bmatrix 0_{mn}\\
\noalign{\medskip}
I_n\endbmatrix.$$
In this case the wave function appearing in
(3.14) is given by
$$\Psi(\lambda,x)=\bmatrix \phi_1(\lambda,x)&\bar\phi_1(\lambda,x)\\
\noalign{\medskip}
\phi_2(\lambda,x)&\bar\phi_2(\lambda,x)\endbmatrix,$$
where
$\bmatrix \phi_1(\lambda,x)\\
\phi_2(\lambda,x)\endbmatrix$ is the $(m+n)\times m$ Jost solution to (5.8) with the
asymptotics
$\bmatrix
e^{-i\lambda x}I_m\\
0_{nm}\endbmatrix$ as $x\to-\infty$
and
$\bmatrix \bar\psi_1(\lambda,x)\\
\bar\psi_2(\lambda,x)\endbmatrix$ is the $(m+n)\times n$ Jost solution to (5.8)
with the asymptotics
$\bmatrix 0_{mn}\\
e^{i\lambda x}I_n\endbmatrix$ as $x\to-\infty.$

In the next three examples we discuss the Schr\"o\-din\-ger equation
on the full and half lines.
In each
example we have $n$ bound states at $k=i\kappa_j$ with the corresponding norming
constant $c_j$. Here $\kappa_1,\dots,\kappa_n$ are distinct positive numbers
and $c_1,\dots,c_n$ are positive. We begin with the familiar example of the
Schr\"o\-din\-ger equation on the full line [15] and then discuss the
Schr\"o\-din\-ger equation on the half line with various boundary conditions
at $x=0$ [5].

\noindent {\bf Example 5.5} Consider the Schr\"o\-din\-ger equation
on the full line
$$-\displaystyle\frac{d^2\varphi}{dx^2}+u(x)\,
\varphi=k^2\varphi,\qquad x\in(-\infty,+\infty),$$
where $u$ is a real-valued integrable potential
with a finite first moment.
The corresponding (left) Marchenko equation is a scalar integral
equation and
is given by (1.7), from whose solution
the potential
$u$ is recovered as
$$u(x)=-2\displaystyle\frac{d \alpha(x,x)}{dx}.$$
In this case, $f(x)$ and $g(y)$ appearing in (1.13) are given [5] by
$$f(x)=\bmatrix c_1 e^{-\kappa_1 x}&\dots&c_n e^{-\kappa_n x}\endbmatrix,\quad
g(y)=\bmatrix
c_1 e^{-\kappa_1 y}\\
\vdots
\\
c_n e^{-\kappa_n y}
\endbmatrix,$$
where $c_j$ is the norming constant for the bound state at $k=i\kappa_j$.
The wave function appearing in (3.14) corresponds to the Jost solution from the left
satisfying $\Psi(k,x)=e^{ikx}[1+o(1)]$ as $x\to+\infty.$
In this case (3.14) holds with the involution $J$ being
the scalar quantity equal to $-1.$

\noindent {\bf Example 5.6} Consider the Schr\"o\-din\-ger equation
on the half line with the Dirichlet boundary condition at the origin, i.e.
$$-\displaystyle\frac{d^2\varphi}{dx^2}+u(x)\,\varphi=k^2\varphi,\quad \varphi(k,0)=0.
\tag 5.15$$
The Gel'fand-Levitan integral equation
arising in the related inverse scattering theory is given by
$$\alpha(x,y)+\omega(x,y)+\int_0^x dy\,\alpha(x,z)\,\omega(z,y)=0,
\qquad 0<y<x,\tag 5.16$$
where the kernel $\omega$ is real valued and symmetric, i.e.
$\omega(x,y)=\omega(y,x),$ and thus it satisfies (1.10).
Note that (5.16) is already in the form of (1.15).
The potential is recovered as
$$u(x)=2\displaystyle\frac{d \alpha(x,x)}{dx}.$$
In this case, $f(x)$ and $g(y)$ appearing in (1.13) are given by given [5] by
$$f(x)=\bmatrix \displaystyle\frac{c_1}{\kappa_1}\sinh(\kappa_1 x)&\dots&
\displaystyle\frac{c_n}{\kappa_n}\sinh(\kappa_n x)\endbmatrix,\quad
g(y)=\bmatrix
\displaystyle\frac{c_1}{\kappa_1}\sinh(\kappa_1 y)\\
\vdots
\\
\displaystyle\frac{c_n}{\kappa_n}\sinh(\kappa_n y)
\endbmatrix,$$
where $c_j$ is the norming constant corresponding the bound state
at $k=i\kappa_j.$ In this case
the wave function $\Psi(k,x)$ is the
regular solution to (5.15) satisfying the
initial conditions $\Psi(k,0)=0$ and
$\Psi'(k,0)=1,$
where the prime denotes the $x$-derivative.
The relationship between
the wave function $\Psi(k,x)$ and $\alpha(x,y),$ instead of (3.14), is given by [5]
$$\Psi(k,x)=\displaystyle\frac{\sin(kx)}{k}+\int_0^x dy\,\alpha(x,y)\,
\displaystyle\frac{\sin(ky)}{k},\qquad 0<y<x,$$
with the inverse transform given by
$$\alpha(x,y)=\displaystyle\frac{1}{\pi}\int_{-\infty}^\infty dk\,k\,\Psi(k,x)\,\sin(k,y),
\qquad 0<y<x.$$

\noindent {\bf Example 5.7} Consider the Schr\"o\-din\-ger equation
on the half line with a selfadjoint boundary condition at the origin, i.e.
$$-\displaystyle\frac{d^2\varphi}{dx^2}+u(x)\,\varphi=k^2\varphi,\quad \varphi'(k,0)
+\cot\theta\,\varphi(k,0)=0,\tag 5.17$$
where $\theta$ is a constant in the interval $(0,\pi).$
The Gel'fand-Levitan integral equation
arising in the related inverse scattering theory is given by
$$\alpha(x,y)+\omega(x,y)+\int_0^x dz\,\alpha(x,z)\,\omega(z,y)=0,
\qquad 0<y<x,\tag 5.18$$
where the kernel $\omega$ is real valued and symmetric, i.e.
$\omega(x,y)=\omega(y,x),$ and thus it satisfies (1.10).
Note that (5.18) is already in the form (1.15).
The potential is recovered as
$$u(x)=2\displaystyle\frac{d \alpha(x,x)}{dx}.$$
In this case, $f(x)$ and $g(y)$ appearing in (1.13) are given [5] by
$$f(x)=\bmatrix c_1\cosh(\kappa_1 x)&\dots&c_n\cosh(\kappa_n x)\endbmatrix,\quad
g(y)=\bmatrix
c_1\cosh(\kappa_1 y)\\
\vdots
\\
c_n\cosh(\kappa_n y)
\endbmatrix,$$
where $c_j$ is the norming constant for the bound state at $k=i\kappa_j$
as in Example 5.6.
In this case the wave function $\Psi(k,x)$ is the
regular solution to (5.17) satisfying the
initial conditions $\Psi(k,0)=1$ and
$\Psi'(k,0)=-\cot\theta.$
The relationship between
$\Psi(k,x)$ and $\alpha(x,y),$ instead of (3.14), is given by
$$\Psi(k,x)=\cos(kx)+\int_0^x dy\,\alpha(x,y)\,
\cos(ky),\qquad 0<y<x,$$
with the inverse transform given by
$$\alpha(x,y)=\displaystyle\frac{1}{\pi}\int_{-\infty}^\infty dk\,\Psi(k,x)\,\cos(ky),
\qquad 0<y<x.$$

\vskip 10 pt
\noindent {\bf 6. DARBOUX TRANSFORM FOR THE ZAKHAROV-SHABAT SYSTEM}
\vskip 3 pt

In order to illustrate the significance of the results presented in this paper,
we will derive the Darboux transformation for the Zakharov-Shabat system given
in (5.1) when one bound state is added to the spectrum.
In particular, to the spectrum of (5.1) we will add one bound state
at $\lambda=\lambda_1\in {\bold C}^+$ with the norming constant $c_1,$
where $c_1$ is a complex constant and we use
${\bold C}^+$ to denote the upper half complex plane.
The potential
appearing in (5.1) will
then change from $u(x)$ to
$\tilde u(x)$ and the wave function appearing in (5.7) will change from $\Psi(\lambda,x)$ to
$\tilde \Psi(\lambda,x).$ With the help of (3.20), (3.21), and (5.5),
we will explicitly evaluate
$\tilde u(x)-u(x)$ and $\tilde \Psi(\lambda,x)-\Psi(\lambda,x)$ in terms of
$u(x),$ $\lambda_1,$ $c_1,$ and $\Psi(\lambda,x).$ Note that
$\tilde u(x)$ and $\tilde \Psi(\lambda,x)$ each consist of a one-parameter family
of potentials and wave functions, respectively, with
$c_1$ being the parameter.
 From (5.7) we see
that, in evaluating
$\tilde \Psi(\lambda,x)-\Psi(\lambda,x),$
it is sufficient to obtain $\bmatrix
\tilde\psi_1(\lambda,x)\\
\tilde\psi_2(\lambda,x)\endbmatrix-
\bmatrix
\psi_1(\lambda,x)\\
\psi_2(\lambda,x)\endbmatrix,$
where
$\bmatrix \psi_1(\lambda,x)\\
\psi_2(\lambda,x)\endbmatrix$ is the Jost solution to (5.1) with the
asymptotics
$\bmatrix 0\\
e^{i\lambda x}\endbmatrix$ as $x\to+\infty.$ At the end of this section we will compare our
result with those in literature.

Before we derive our Darboux transformation, we need some identities related to
the solutions to (5.1). Let us use an overdot to indicate the derivative
with respect to $\lambda.$ The following result is already known and its proof is omitted.
Its proof can easily be obtained by using the integral representation
of the Jost solution [30].

\noindent {\bf Proposition 6.1} {\it Let $\bmatrix \psi_1(\lambda,x)\\
\psi_2(\lambda,x)\endbmatrix$ be the Jost solution to (5.1) with the
asymptotics
$\bmatrix 0\\
e^{i\lambda x}\endbmatrix$ as $x\to+\infty,$ where $u$ is an integrable potential.
If $\lambda_1\in{\bold C}^+,$ then
$\psi_1(\lambda_1,x),$
$\psi_2(\lambda_1,x),$
$\dot \psi_1(\lambda_1,x),$ and $\dot \psi_2(\lambda_1,x)$
all vanish as $x\to+\infty.$}

Even though the identities in the following
proposition are known [1-3,30,35], we provide a brief proof for the convenience of the reader.

\noindent {\bf Proposition 6.2} {\it Let $\bmatrix \psi_1(\lambda,x)\\
\psi_2(\lambda,x)\endbmatrix$ be the Jost solution to (5.1) with the
asymptotics
$\bmatrix 0\\
e^{i\lambda x}\endbmatrix$ as $x\to+\infty,$ where $u$ is an integrable potential.
We then have the following identities:}
$$\displaystyle\frac{d}{dx}\left| \matrix
\psi_1(\lambda,x)&-\psi_2(\lambda,x)^\ast
\\
\noalign{\medskip}
\psi_2(\lambda,x)&\psi_1(\lambda,x)^\ast
\endmatrix\right|
=2\,\text{Im}[\lambda] \,\left[
|\psi_1(\lambda,x)|^2-|\psi_2(\lambda,x)|^2\right],\tag 6.1$$
$$\displaystyle\frac{d}{dx}\left| \matrix
\psi_1(\lambda,x)&\dot\psi_1(\lambda,x)
\\
\noalign{\medskip}
\psi_2(\lambda,x)&\dot\psi_2(\lambda,x)
\endmatrix\right|=2i\,
\psi_1(\lambda,x)\,\psi_2(\lambda,x),\tag 6.2$$
$$\displaystyle\frac{d}{dx}\left| \matrix
\psi_1(\lambda,x)&\psi_1(\lambda_1,x)
\\
\noalign{\medskip}
\psi_2(\lambda,x)&\psi_2(\lambda_1,x)
\endmatrix\right|=-i(\lambda-\lambda_1)[
\psi_1(\lambda,x)\,
\psi_2(\lambda_1,x)
+\psi_2(\lambda,x)
\psi_1(\lambda_1,x)],\tag 6.3$$
$$\displaystyle\frac{d}{dx}\left| \matrix
\psi_1(\lambda,x)&-\psi_2(\lambda_1,x)^\ast
\\
\noalign{\medskip}
\psi_2(\lambda,x)&\psi_1(\lambda_1,x)^\ast
\endmatrix\right|=-i(\lambda-\lambda_1^\ast)[
\psi_1(\lambda,x)\,
\psi_1(\lambda_1,x)^\ast
-\psi_2(\lambda,x)
\psi_2(\lambda_1,x)^\ast],\tag 6.4$$
{\it where $\text{Im}[\lambda]$
is used to denote the imaginary part of $\lambda.$}

\noindent PROOF: From (5.1) we obtain
$$\cases
\psi_1'(\lambda,x)^\ast=i\lambda^\ast\psi_1(\lambda,x)^\ast+u(x)^\ast\,\psi_2(\lambda,x)^\ast,\\
\noalign{\medskip}
\psi_2'(\lambda,x)^\ast=-i\lambda^\ast\psi_2(\lambda,x)^\ast-u(x)\,\psi_1(\lambda,x)^\ast,\endcases
\tag 6.5$$
$$\cases
\dot\psi_1'(\lambda,x)=-i\psi_1(\lambda,x)
-i\lambda\dot \psi_1(\lambda,x)+u(x)\,\dot\psi_2(\lambda,x),\\
\noalign{\medskip}
\dot\psi_2'(\lambda,x)=i
\psi_2(\lambda,x)+i\lambda\dot\psi_2(\lambda,x)-u(x)^\ast\,\dot\psi_1(\lambda,x).\endcases
\tag 6.6$$
The identities given in
(6.1)-(6.4) are derived directly from (5.1), (6.5), and (6.6). \qed

Using Propositions 6.1 and 6.2, we have the following result.

\noindent {\bf Corollary 6.3} {\it Let $\bmatrix \psi_1(\lambda,x)\\
\psi_2(\lambda,x)\endbmatrix$ be the Jost solution to (5.1) with the
asymptotics
$\bmatrix 0\\
e^{i\lambda x}\endbmatrix$ as $x\to+\infty,$ where $u$ is an integrable potential.
We then have the following identities:}
$$\int_x^\infty ds\,
\psi_1(\lambda_1,s)\,\psi_2(\lambda_1,s)=
\displaystyle\frac{i}{2}\left[
\psi_1(\lambda_1,x)\,\dot\psi_2(\lambda_1,x)-
\psi_2(\lambda_1,x)\,\dot\psi_1(\lambda_1,x)
\right],
\tag 6.7$$
$$\int_x^\infty ds \left(|\psi_1(\lambda_1,s)|^2
-|\psi_2(\lambda_1,s)|^2\right)=-
\displaystyle\frac{|\psi_1(\lambda_1,x)|^2+|\psi_2(\lambda_1,x)|^2}{2\,\text{Im}[\lambda_1]},
\tag 6.8$$
$$\aligned
\int_x^\infty
ds\,[\psi_2(\lambda_1,s)\,\psi_1(\lambda,s)+&\psi_1(\lambda_1,s)\,\psi_2(\lambda,s)]
\\
&=\displaystyle\frac{i}{\lambda-\lambda_1}\left[\psi_1(\lambda_1,x)\,
\psi_2(\lambda,x)-
\psi_2(\lambda_1,x)\,\psi_1(\lambda,x)
\right]
,\endaligned\tag 6.9$$
$$\aligned
\int_x^\infty
ds\,[\psi_1(\lambda_1,s)^\ast\,\psi_1(\lambda,s)
&-\psi_2(\lambda_1,s)^\ast\,\psi_2(\lambda,s)]
\\
&=\displaystyle\frac{-i}{\lambda-\lambda_1^\ast}\left[
\psi_1(\lambda_1,x)^\ast\,\psi_1(\lambda,x)
+\psi_2(\lambda_1,x)^\ast\,\psi_2(\lambda,x)
\right]
.\endaligned\tag 6.10$$

Let us now use our systematic approach to derive
the one-parameter family of the Darboux transformations for (5.1)
when one bound state at $\lambda=\lambda_1\in {\bold C}^+$ with the norming constant $c_1$
is added to the spectrum.
Let us choose in (5.6)
$$A_{\text l}=-i\lambda_1,\quad B_{\text l}=1,\quad C_{\text l}=c_1.$$
Then, using (5.6) and (5.7) in (4.5), (4.7), and (4.10) we obtain
$$n(x)=\bmatrix c_1\psi_1(\lambda_1,x)&\psi_2(\lambda_1,x)^*\\ \noalign{\medskip}
c_1\psi_2(\lambda_1,x)&-\psi_1(\lambda_1,x)^*\endbmatrix,\tag 6.11$$
$$q(x)=\bmatrix \psi_2(\lambda_1,x)&\psi_1(\lambda_1,x)\\ \noalign{\medskip}
c_1^*\psi_1(\lambda_1,x)^*&-c_1^*\psi_2(\lambda_1,x)^*\endbmatrix,\tag 6.12$$
$$\Gamma(x)=\bmatrix \Gamma_1(x)&\Gamma_2(x)\\
\noalign{\medskip}
\Gamma_3(x)&\Gamma_4(x)\endbmatrix,\tag 6.13$$
where we have defined
$$\Gamma_1(x):=\Gamma_4(x)^*:=1+2c_1\int_x^\infty ds\,
\psi_1(\lambda_1,s)\,\psi_2(\lambda_1,s),\tag 6.14$$
$$\Gamma_2(x):=
-\displaystyle\frac{\Gamma_3(x)}{|c_1|^2}:=
-\int_x^\infty ds\,\left(|\psi_1(\lambda_1,s)|^2
-|\psi_2(\lambda_1,s)|^2\right).\tag 6.15$$
Using (6.7) in (6.14) and using (6.8) in (6.15) we get
$$\Gamma_4(x)^\ast=\Gamma_1(x)=1+ic_1\left[
\psi_1(\lambda_1,x)\,\dot\psi_2(\lambda_1,x)-
\psi_2(\lambda_1,x)\,\dot\psi_1(\lambda_1,x)
\right],\tag 6.16$$
$$-\displaystyle\frac{\Gamma_3(x)}{|c_1|^2}=\Gamma_2(x)=
\displaystyle\frac{|\psi_1(\lambda_1,x)|^2+|\psi_2(\lambda_1,x)|^2}{2\,\text{Im}[\lambda_1]}.
\tag 6.17$$

Next, we present the main result in this section, namely, the Darboux transformation
for the Zakharov-Shabat system given in (5.1).

\noindent {\bf Theorem 6.4} {\it When one bound state at
$\lambda=\lambda_1\in {\bold C}^+$ with the norming constant $c_1$
is added to the spectrum of (5.1), the potential
$u(x)$ changes to
$\tilde u(x)$ and
the Jost solution
$\bmatrix \psi_1(\lambda,x)\\
\psi_2(\lambda,x)\endbmatrix$ changes
to $\bmatrix \tilde\psi_1(\lambda,x)\\
\tilde\psi_2(\lambda,x)\endbmatrix$
according to the following Darboux transformation:}
$$\tilde u(x)-u(x)=
\displaystyle\frac{P_0(x)}{|\Gamma_1(x)|^2+|c_1|^2 \Gamma_2(x)^2},\tag 6.18$$
$$\bmatrix
\tilde\psi_1(\lambda,x)\\
\noalign{\medskip}
\tilde\psi_2(\lambda,x)\endbmatrix-
\bmatrix
\psi_1(\lambda,x)\\
\noalign{\medskip}
\psi_2(\lambda,x)\endbmatrix=\displaystyle\frac{1}{|\Gamma_1(x)|^2+|c_1|^2\Gamma_2(x)^2}
\bmatrix P_1&P_2\\
\noalign{\medskip}
P_3&P_4\endbmatrix
\bmatrix P_5\\
\noalign{\medskip}
P_6\endbmatrix,\tag 6.19$$
{\it
where we have $\Gamma_1(x)$ and $\Gamma_2(x)$ are as in (6.16) and
(6.17), respectively, and}
$$\aligned P_0:=&2c_1\psi_1(\lambda_1,x)^2
\Gamma_1(x)^\ast-2c_1^\ast[\psi_2(\lambda_1,x)^\ast]^2
\Gamma_1(x)\\
&+4|c_1|^2\psi_1(\lambda_1,x)\,
\psi_2(\lambda_1,x)^\ast\, \Gamma_2(x),\endaligned$$
$$P_1:=-|c_1|^2\,\psi_2(\lambda_1,x)^\ast
\,\Gamma_2(x)-c_1\psi_1(\lambda_1,x)\,\Gamma_1(x)^\ast,$$
$$P_2:=|c_1|^2\,\psi_1(\lambda_1,x)
\,\Gamma_2(x)-c_1^\ast\,\psi_2(\lambda_1,x)^\ast\,\Gamma_1(x),$$
$$P_3:=|c_1|^2\,\psi_1(\lambda_1,x)^\ast
\,\Gamma_2(x)-c_1\psi_2(\lambda_1,x)\,\Gamma_1(x)^\ast,$$
$$P_4:=|c_1|^2\,\psi_2(\lambda_1,x)
\,\Gamma_2(x)+c_1^\ast\,\psi_1(\lambda_1,x)^\ast\,\Gamma_1(x),$$
$$P_5:=\displaystyle\frac{i}{\lambda-\lambda_1}\left[
\psi_1(\lambda_1,x)\,\psi_2(\lambda,x)-
\psi_2(\lambda_1,x)\,\psi_1(\lambda,x)
\right],$$
$$P_6:=\displaystyle\frac{-i}{\lambda-\lambda_1^\ast}\left[
\psi_1(\lambda_1,x)^\ast\,\psi_1(\lambda,x)
+\psi_2(\lambda_1,x)^\ast\,\psi_2(\lambda,x)\,
\right].$$

\noindent PROOF: From (3.21) and (5.5)
we have
$$\tilde u(x)-u(x)=2\bmatrix 1&0\endbmatrix n(x)\,\Gamma(x)^{-1}q(x)
\bmatrix 0\\
\noalign{\medskip}
1\endbmatrix.\tag 6.20$$
Thus, using (6.11)-(6.13), (6.16), and (6.17) in (6.20) we obtain (6.18).
 From (3.18), (3.20), (5.7), and (6.12)
it follows that
$$\bmatrix
\tilde\psi_1(\lambda,x)\\
\noalign{\medskip}
\tilde\psi_2(\lambda,x)\endbmatrix-
\bmatrix
\psi_1(\lambda,x)\\
\noalign{\medskip}
\psi_2(\lambda,x)\endbmatrix=-
n(x)\,\Gamma(x)^{-1}\int_x^\infty ds\,q(s)
\bmatrix
\psi_1(\lambda,s)\\
\noalign{\medskip}
\psi_2(\lambda,s)\endbmatrix.\tag 6.21$$
Using (6.11)-(6.13), (6.16), and (6.17) on the right
hand side of (6.21), with the help of
(6.9) and (6.10) we obtain (6.19). \qed

Let us now compare our systematic approach to the Darboux transformation
for the Zakharov-Shabat system given in
Theorem 6.4 with some of the results
in the literature.
Some special cases of the Darboux transformations were obtained explicitly
[10,23,24] for the Zakharov-Shabat system when a bound state is added
at $\lambda=\lambda_1,$ whereas our own method provides all the Darboux
transformations in this case.
When a bound state is added
at $\lambda=\lambda_1,$
one special Darboux transformation,
named an elementary Darboux transform [10,23,24], corresponds to (cf. (1.9) of [10])
$$\tilde u(x)=\displaystyle\frac{\psi_2(\lambda_1,x)^*}{\psi_1(\lambda_1,x)^*}
=u'(x)+2i\lambda_1 u(x)-\displaystyle\frac{\psi_2(\lambda_1,x)}
{\psi_1(\lambda_1,x)}\,u(x)^2.
\tag 6.22$$
We will see that this Darboux transformation
corresponds to the choice of $1/c_1=0$ in our one-parameter family given in (6.18).
A second elementary Darboux transform  [10,23,24] is given by (cf. (1.11) of [10])
$$\tilde u(x)=\displaystyle\frac{\psi_1(\lambda_1,x)}{\psi_2(\lambda_1,x)}
=-u'(x)-2i\lambda_1^*\, u(x)-\displaystyle\frac{\psi_1(\lambda_1,x)^*}
{\psi_2(\lambda_1,x)^*}\,u(x)^2.\tag 6.23$$
which corresponds to an added bound state at $\lambda=\lambda_1^\ast.$
A combination of these two yields the following Darboux transformation (cf. (1.14) of [10])
when a bound state at $\lambda=\lambda_1$ and another one at
$\lambda=\lambda_1^\ast$ are added to the spectrum of (5.1):
$$\tilde u(x)-u(x)=
\displaystyle\frac{4\,\text{Im}[\lambda_1]\,\psi_1(\lambda_1,x)\,\psi_2(\lambda_1,x)^*}
{|\psi_1(\lambda_1,x)|^2+|\psi_2(\lambda_1,x)|^2}.\tag 6.24$$
The Jost solution is then transformed according to (cf. (2.41) of [10])
$$\bmatrix\tilde\psi_1(\lambda,x)
\\
\noalign{\medskip}
\tilde\psi_2(\lambda,x)\endbmatrix=
\left(
-\displaystyle\frac i2(\lambda-\lambda_1)+
\displaystyle\frac
{\text{Im}[\lambda_1]\,S(\lambda_1,x)}
{|\psi_1(\lambda_1,x)|^2+|\psi_2(\lambda_1,x)|^2}
\right)
\bmatrix \psi_1(\lambda,x)
\\
\noalign{\medskip}
 \psi_2(\lambda,x)\endbmatrix,\tag 6.25$$
where the matrix $S(\lambda_1,x)$ is defined as
$$S(\lambda_1,x):=
\bmatrix |\psi_2(\lambda_1,x)|^2&-
\psi_1(\lambda_1,x)\,\psi_2(\lambda_1,x)^*
\\
\noalign{\medskip}
-
\psi_1(\lambda_1,x)^*\,\psi_2(\lambda_1,x)
& |\psi_1(\lambda_1,x)|^2\endbmatrix.$$
Note that (6.25) implies that
$$\tilde\psi_1(\lambda_1,x)=\tilde\psi_2(\lambda_1,x)=0.$$
The norming constant $c_1$ appearing in (6.18) is given by
$$c_1=\displaystyle\frac{-1}{2\displaystyle\int_{-\infty}^\infty ds\,
\tilde \psi_1(\lambda_1,s)\,\tilde \psi_2(\lambda_1,s)},$$
and hence the Darboux transformation given in (6.22)
corresponds to a very particular choice of the norming
constant $c_1,$ namely $1/c_1=0.$ Similarly, for
the Darboux transformation given in (6.23), the reciprocal
of the
norming constant is zero. Consequently,
the Darboux transformation given in (6.24) corresponds to one particular choice
among the two-parameter family $\tilde u(x).$

We can also see that each of the three Darboux transforms given in (6.22)-(6.24) are very
special by considering
the easiest case where $u(x)=0.$ In that case (6.22) is understood
as $\tilde u(x)=0$ and (6.23) and (6.24) each impose $\tilde u(x)=0.$
When $u(x)=0$, the Jost solution $\bmatrix \psi_1(\lambda,x)\\
\psi_2(\lambda,x)\endbmatrix$ to (5.1) with the
asymptotics
$\bmatrix 0\\
e^{i\lambda x}\endbmatrix$ as $x\to+\infty$
is given by
$$\psi_1(\lambda,x)=0,\quad \psi_2(\lambda,x)=e^{i\lambda x}.\tag 6.26$$
Then, we see that (6.24) also imposes $\tilde u(x)=0.$
On the other hand, using (6.26) in
(6.11), (6.12), (6.16), (6.17), and Theorem 6.4, our own procedure
yields the one-parameter family of potentials and wave functions
$$\tilde u(x)=\displaystyle\frac{-8c_1^\ast\,
\left(\text{Im}[\lambda_1]\right)^2
e^{-2i\lambda_1^\ast x}}
{4\left(\text{Im}[\lambda_1]\right)^2+|c_1|^2
e^{-4\text{Im}[\lambda_1]\, x}},\tag 6.27$$
$$\tilde\psi_1(\lambda,x)=
\displaystyle\frac{4ic_1^\ast\,
\left(\text{Im}[\lambda_1]\right)^2
e^{-2i\lambda_1^\ast x+i\lambda x}}
{\left(\lambda-\lambda_1^\ast\right)\left[4\left(\text{Im}[\lambda_1]\right)^2+|c_1|^2
e^{-4\text{Im}[\lambda_1]\, x}\right]},$$
$$\tilde\psi_2(\lambda,x)=e^{i\lambda x}-
\displaystyle\frac{2i|c_1|^2\,
\left(\text{Im}[\lambda_1]\right)
e^{-4\text{Im}[\lambda_1]\, x+i\lambda x}}
{\left(\lambda-\lambda_1^\ast\right)\left[4\left(\text{Im}[\lambda_1]\right)^2+|c_1|^2
e^{-4\text{Im}[\lambda_1]\, x}\right]}.$$
Using [1-3,29]
$$\tilde\psi_2(\lambda,x)=e^{i\lambda x}
\left[\displaystyle\frac{1}{\tilde T(\lambda)}+o(1)\right],\qquad
x\to-\infty,$$
where $\tilde T(\lambda)$ is the transmission coefficient corresponding
to $\tilde u(x),$ we get $\tilde T(\lambda)=
\displaystyle\frac{\lambda-\lambda_1^\ast}{\lambda-\lambda_1}.$
We note that the quantity given in (6.27) is the one-soliton
potential and the corresponding reflection coefficients are zero.

In [29] a Darboux transformation was studied for the Zakharov-Shabat system in two
spatial variables, and by eliminating one of the variables, the corresponding
Darboux transformation was provided in the form of
an ordinary differential equation (cf. (4.9a) of [29]) involving
$\tilde u'(x),$ $u'(x),$ $\tilde u(x),$ and $u(x).$ The same differential
equation
was derived earlier (cf. (4.7) of [26]) by Levi, Ragnisco, and Sym, who
studied the equivalence of the dressing method [35]
and the Darboux transformation for the Schr\"odinger equation and indicated
that the result also holds for the
Zakharov-Shabat system. These authors
obtained a
formula, which is given as (4.6) in their paper [26], connecting
$\tilde u(x)$ and $u(x)$ through
an intermediate function $\bar \xi$ expressed in terms of the four entries of
a matrix-valued solution to (5.1).
A first-order differential equation similar to (4.9a) of [29] was
earlier derived by Gerdzhikov and Kulish (cf. (14) of [19]).
In [28], when $N$ bound states
are added to the spectrum, the change in the potential
is expressed (cf (20) of [28]) in terms of ratios of the determinants
of two $2N\times 2N$ matrices differing only in their last columns;
such matrices are constructed by determining the zeros of certain
polynomial equations in $\lambda$ and by solving
certain linear algebraic equations.
Various other authors (see, e.g. [22]) presented similar formulas for the change
in the potential when bound states are added to the spectrum. In most of these
papers a ``Darboux matrix"
is constructed
connecting the wave functions of the original and perturbed problems.

One criticism of the result
of [26] is that a matrix solution to (5.1) was evaluated (cf. (3.6) of [26])
at a $\lambda$-value
on the upper-half complex plane and also evaluated
(cf. (3.8) of [26]) at a $\lambda$-value
on the lower-half complex plane. The same concern also applies
to other works (see, e.g. (19) of [28] and (3.8) of [22]). In general, we cannot expect
the entries of a matrix solution to
(5.1) to have extensions in $\lambda$ to both upper and lower complex planes, unless
the class of potentials
$u(x)$ is very restrictive.

\vskip 10 pt
\noindent {\bf A. APPENDIX: SOME NORM ESTIMATES}
\vskip 3 pt

In this appendix we derive some boundedness properties of the integral operators
$\Omega$ defined in Section 1.

Suppose $\omega(x,y)$ is an $N\times N$ matrix function satisfying (1.2). Let
$$N_1(x):=\sup_{y>x}\,\int_x^\infty dz\,\|\omega(y,z)\|,\quad
N_\infty(x):=\sup_{y>x}\,\int_x^\infty dz\,\|\omega(z,y)\|,$$
so that (1.2) amounts to $N_1(x)+N_\infty(x)<+\infty$.

\noindent{\bf Proposition A.1} {\it If the assumption in (1.2) is satisfied,
then the integral
operator $\Omega$ appearing in (1.3) is bounded on ${\Cal H}_1^{M\times N}$ with norm bound
$N_1(x)$ and on ${\Cal H}_\infty^{M\times N}$ with norm bound $N_\infty(x)$.}

\noindent PROOF: We directly verify that
$$\int_x^\infty dy\,\|(\beta\Omega)(x,y)\|
\le\int_x^\infty dy\int_x^\infty dz\,\|\beta(x,z)\|\,\|\omega(z,y)\|
\le N_1(x)\int_x^\infty dz\,\|\beta(x,z)\|,$$
and
$$\|(\beta\Omega)(x,y)\|\le N_\infty(x)\sup_{y>x}\|\beta(x,y)\|,$$
which proves the proposition. \qed

It is now clear that the (scalar) integral operator $L_\omega$ defined by
$$(L_\omega h)(y):=\int_x^\infty dz\,h(z)\|\omega(z,y)\|,$$
is bounded on $L^1(x,+\infty)$ with norm bound $N_1(x)$ and on
$L^\infty(x,+\infty)$ with norm bound $N_\infty(x)$. By the Riesz-Thorin
interpolation theorem (cf. [36], Vol. II, Sec. XII.1), $L_\omega$ is bounded
on $L^p(x,+\infty)$ for $p\in(1,+\infty)$ with norm bounded above by
$$N_1(x)^{1/p}N_\infty(x)^{1-(1/p)}.$$
Since $\Omega$ maps a dense linear subspace of
${\Cal H}_p^{M\times N}$ (namely, its intersection with
${\Cal H}_1^{M\times N}\cap {\Cal H}_\infty^{M\times N}$) into
${\Cal H}_1^{M\times N}\cap {\Cal H}_\infty^{M\times N}$,
with $h(z)=\|\beta(x,z)\|$ we obtain the estimate
$$\aligned\left[\int_x^\infty dz\,\|(\beta\Omega)(x,z)\|^p\right]^{1/p}
&\le\|L_\omega h\|_p\le N_1(x)^{1/p}N_\infty(x)^{1-(1/p)}\|h\|_p\\
&=N_1(x)^{1/p}N_\infty(x)^{1-(1/p)}\left[\int_x^\infty dz\,
\|\beta(x,z)\|^p\right]^{1/p},\endaligned$$
where $||\cdot||_p$ is the $L^p$-norm.
Hence, $\Omega$ is bounded on ${\Cal H}_p^{M\times N}$
for $p\in(1,+\infty)$ as well.

\vskip 10 pt

\noindent{\bf Acknowledgment}. The research leading to this article was
supported in part by the National Science Foundation under grant
DMS-0610494 and by INdAM-GNCS.

\vskip 10 pt

\noindent {\bf REFERENCES}

\item{[1]} M. J. Ablowitz and P. A. Clarkson, {\it Solitons, nonlinear
evolution equations and inverse scattering,} Cambridge Univ. Press, Cambridge,
1991.

\item{[2]} M. J. Ablowitz, D. J. Kaup, A. C. Newell, and H. Segur, {\it The
inverse scattering transform-Fourier analysis for nonlinear problems,} Stud.
Appl. Math. {\bf 53}, 249--315 (1974).

\item{[3]} M. J. Ablowitz and H. Segur, {\it
Solitons and the inverse scattering transform,} SIAM, Philadelphia, 1981.

\item{[4]} T. Aktosun, F. Demontis, and C. van der Mee,
{\it Exact solutions to the focusing nonlinear Schr\"o\-din\-ger equation,}
Inverse Problems {\bf 23}, 2171--2195 (2007).

\item{[5]} T. Aktosun and R. Weder, {\it Inverse spectral-scattering problem
with two sets of discrete spectra for the radial Schr\"o\-din\-ger equation,}
Inverse Problems {\bf 22}, 89--114 (2006).

\item{[6]} V. A. Ambarzumian, {\it On the scattering of light by planetary
atmospheres}, Astr. J. Soviet Union {\bf 19}, 30--41 (1942).

\item{[7]} I. W. Busbridge, {\it The mathematics of radiative transfer,} Cambridge Univ.
Press, London, 1960.

\item{[8]} F. Calogero and A. Degasperis, {\it Nonlinear evolution
equations solvable by the inverse spectral transform. I,}
Nuovo Cimento B {\bf 32}, 201--242 (1976).

\item{[9]} F. Calogero and A. Degasperis, {\it Spectral transform and solitons,}
North-Holland, Amsterdam, 1982.

\item{[10]} R. C. Cascaval, F. Gesztesy, H. Holden, and Y. Latushkin,
{\it Spectral analysis of Darboux transformations for the focusing NLS
hierarchy,} J. Anal. Math. {\bf 93}, 139--197 (2004).

\item{[11]} S. Chandrasekhar,
{\it Radiative transfer,}
Dover Publ., New York, 1960.

\item{[12]} S. Chandrasekhar, {\it The transfer of
radiation in stellar atmospheres,} Bull. Amer. Math. Soc.
{\bf 53}, 641--711 (1947).

\item{[13]} Hsing-Hen Chen, {\it General derivation of B\"acklund transformations from
inverse scattering problems,} Quart.
Phys. Rev. Lett. {\bf 33}, 925--928 (1974).

\item{[14]} M. M. Crum, {\it Associated Sturm-Liouville systems,} Quart.
J. Math. Oxford Ser. (2) {\bf 8}, 121--127 (1955).

\item{[15]} P. Deift and E. Trubowitz, {\it Inverse scattering on the
line,} Commun. Pure Appl. Math. {\bf 32}, 121--251 (1979).

\item{[16]} F. Demontis, {\it Direct and inverse scattering of the matrix
Zakharov-Shabat system,} Ph.D. thesis, University of Cagliari, Italy, 2007.

\item{[17]} F. Demontis and C. van der Mee, {\it Marchenko
equations and norming constants of the matrix Zakharov-Shabat system,}
Oper. Matrices {\bf 2}, 79--113 (2008).

\item{[18]} L. P. Eisenhart,
{\it A treatise on the differential geometry of curves and surfaces,}
Dover Publ., New York, 1960.

\item{[19]} V. S. Gerdzhikov and P. P. Kulish, {\it Derivation of the B\"acklund
transformation in the formalism of
the inverse scattering problem,} Theor. Math. Phys. {\bf 39}, 327--331
(1979).

\item{[20]} I. C. Gohberg and I. A. Feldman, {\it Convolution equations and
projection methods for their solution,} Transl. Math. Monographs {\bf 41},
Amer. Math. Soc., Providence, RI, 1974.

\item{[21]} I. C. Gohberg and M. G. Krein, {\it Systems of integral equations
on a half-line with kernels depending on the difference of arguments,} Amer.
Math. Soc. Transl. (Ser. 2), {\bf 14}, 217--287 (1960).

\item{[22]} Chao Hao Gu and Zi Xiang Zhou, {\it On Darboux matrices of
B\"acklund transformations for AKNS systems,} Lett. Math. Phys. {\bf 13}, 179--187
(1987).

\item{[23]} B. G. Konopelchenko, {\it Elementary B\"acklund transformations,
nonlinear superposition principle and solutions of the integrable equations,}
Phys. Lett. A {\bf 87}, 445--448 (1982).

\item{[24]} B. G. Konopelchenko and C. Rogers, {\it B\"acklund and reciprocal
transformations: gauge connections,} In: W. F. Ames and C. Rogers (eds.),
{\it Nonlinear equations in the applied sciences,} Academic Press, Boston,
1992, pp. 317--362.

\item{[25]} G. L. Lamb, Jr., {\it B\"acklund transformations for
certain nonlinear evolution equations,}
J. Math. Phys. {\bf 15}, 2157--2165 (1974).

\item{[26]} D. Levi, O. Ragnisco, and A. Sym, {\it Dressing method vs. classical Darboux
transformation,} Nuovo Cimento B {\bf 83}, 34--42 (1984).

\item{[27]} V. B. Matveev and M. A. Salle, {\it Darboux transformations and
solitons,} Springer, Berlin, 1991.

\item{[28]} G. Neugebauer and R. Meinel, {\it General $N$-soliton solution of
the AKNS class on arbitrary background,} Phys. Lett. A {\bf 100}, 467--470
(1984).

\item{[29]} J. J. C. Nimmo, {\it Darboux transformations for a two-dimensional
Zakharov-Shabat /AKNS spectral problem,} Inverse Problems
{\bf 8}, 219--243 (1992).

\item{[30]} S. Novikov, S. V. Manakov, L. P. Pitaevskii, and V. E. Zakharov,
{\it Theory of solitons,} Consultants Bureau, New York, 1984.

\item{[31]} C. Rogers and W. K. Schief, {\it B\"acklund and Darboux
transformations: geometry and modern applications in soliton theory,} Cambridge
University Press, Cambridge, 2002.

\item{[32]} V. V. Sobolev,
{\it A treatise on radiative
transfer}, Van Nostrand, Princeton, New Jersey, 1963.

\item{[33]} V. V. Sobolev,
{\it Light
scattering in planetary atmospheres,} Pergamon Press, Oxford, 1975.

\item{[34]} M. Wadati, H. Sanuki, and K. Konno, {\it Relationships among inverse method, B\"acklund
transformation and an infinite number of conservation laws,} Prog. Theor. Phys. {\bf 53}, 419--436
(1975).

\item{[35]} V. E. Zakharov and A. B. Shabat, {\it A scheme for integrating the
nonlinear equations of mathematical physics by the method of the inverse
scattering transform,} Funct. Anal. App. {\bf 8}, 226--235 (1974).

\item{[36]} A. Zygmund, {\it Trigonometric series}, Vols. I and II, Cambridge
Univ. Press, Cambridge, 1959.

\end